\documentclass[sigconf]{acmart}

\renewcommand\footnotetextcopyrightpermission[1]{} % removes footnote with conference information in first column
\usepackage{amsmath,amssymb,amsfonts}
\usepackage{algorithm,algorithmicx,algpseudocode}
\usepackage{graphicx}
\usepackage{textcomp}
\usepackage{xcolor}
\usepackage{array}
\usepackage{multirow}
\usepackage{longtable}
\usepackage{rotating}
\usepackage{verbatim}
\AtBeginDocument{%
  }

\copyrightyear{2025}
\acmYear{2025}
\setcopyright{rightsretained}
\acmConference[ICPP '25]{54th International Conference on Parallel Processing}{September 08--11, 2025}{San Diego, CA, USA}
\acmBooktitle{54th International Conference on Parallel Processing (ICPP '25), September 08--11, 2025, San Diego, CA, USA}
\acmPrice{}
\acmDOI{10.1145/3754598.3754629}
\acmISBN{979-8-4007-2074-1/25/09}

\begin{document}

%\title{FLEX: Leveraging FPGA-CPU Synergy for VLSI Legalization Acceleration with Efficient Cell Shifting Algorithm}
\title{FLEX: Leveraging FPGA-CPU Synergy for Mixed-Cell-Height Legalization Acceleration}

\author{Xingyu Liu}
\email{xliugu@connect.ust.hk}
\orcid{0009-0006-0897-8670}
\affiliation{%
  \institution{The Hong Kong University of Science and Technology}
  \state{Kowloon}
  \country{Hong Kong}
}

\author{Jiawei Liang}
\email{jliangbr@connect.ust.hk}
\orcid{0000-0003-0748-3183}
\affiliation{%
  \institution{The Hong Kong University of Science and Technology}
  \state{Kowloon}
  \country{Hong Kong}
}

\author{Linfeng Du}
\email{linfeng.du@connect.ust.hk}
\orcid{0000-0002-3007-4890}
\affiliation{%
  \institution{The Hong Kong University of Science and Technology}
  \state{Kowloon}
  \country{Hong Kong}
}

\author{Yipu Zhang}
\email{yzhangqg@connect.ust.hk}
\orcid{0009-0007-9465-5807}
\affiliation{%
  \institution{The Hong Kong University of Science and Technology}
  \state{Kowloon}
  \country{Hong Kong}
}

\author{Chaofang Ma}
\email{cmaaw@connect.ust.hk}
\orcid{0009-0009-2082-4916}
\affiliation{%
  \institution{The Hong Kong University of Science and Technology}
  \state{Kowloon}
  \country{Hong Kong}
}

\author{Hanwei Fan}
\email{hfanah@connect.ust.hk}
\orcid{0000-0002-1177-2108}
\affiliation{%
  \institution{The Hong Kong University of Science and Technology}
  \state{Kowloon}
  \country{Hong Kong}
}

\author{Jiang Xu}
\email{jiang.xu@hkust-gz.edu.cn}
\orcid{0000-0001-9089-7752}
\affiliation{%
  \institution{The Hong Kong University of Science and Technology (Guangzhou)}
  \city{Guangzhou}
  \country{China}
}

\author{Wei Zhang}
%\authornote{Corresponding Author.}
\email{wei.zhang@ust.hk}
\orcid{0000-0002-7622-6714}
\affiliation{%
  \institution{The Hong Kong University of Science and Technology}
  \state{Kowloon}
  \country{Hong Kong}
}


\begin{abstract}
Legalization is a critical yet time-consuming step in very large-scale integration (VLSI) design, tasked with iteratively relocating standard cells to eliminate overlaps while resolving design rule violations. This process is applied after global placement and repeatedly invoked during VLSI physical design. However, increasing spatial constraints and complex design rules impose significant challenges on existing CPU- and GPU-based legalizers, including suboptimal task assignment, inefficient algorithm, and long hardware idle time caused by processing tasks with irregular computational patterns in parallel. 

In this work, we present FLEX, an FPGA-CPU accelerator for mixed-cell-height legalization tasks. We address the above challenges from the following perspectives. First, we optimize the task assignment strategy and perform an efficient task partition between FPGA and CPU to exploit their complementary strengths. Second, a multi-granularity pipelining technique is employed to accelerate the most time-consuming step, finding optimal placement position (FOP), in legalization. At last, we particularly target the computationally intensive cell shifting process in FOP, optimizing the design to align it seamlessly with the multi-granularity pipelining framework for further speedup. Experimental results show that FLEX achieves up to 18.3$\times$ and 5.4$\times$ speedups compared to state-of-the-art CPU-GPU and multi-threaded CPU legalizers with better scalability, while improving legalization quality by 4\% and 1\%.
\end{abstract}




\maketitle

%\vspace{-1pt}
\section{INTRODUCTION}

\begin{figure} [h]
\centering
\vspace{-8pt}
\includegraphics[width=72mm]{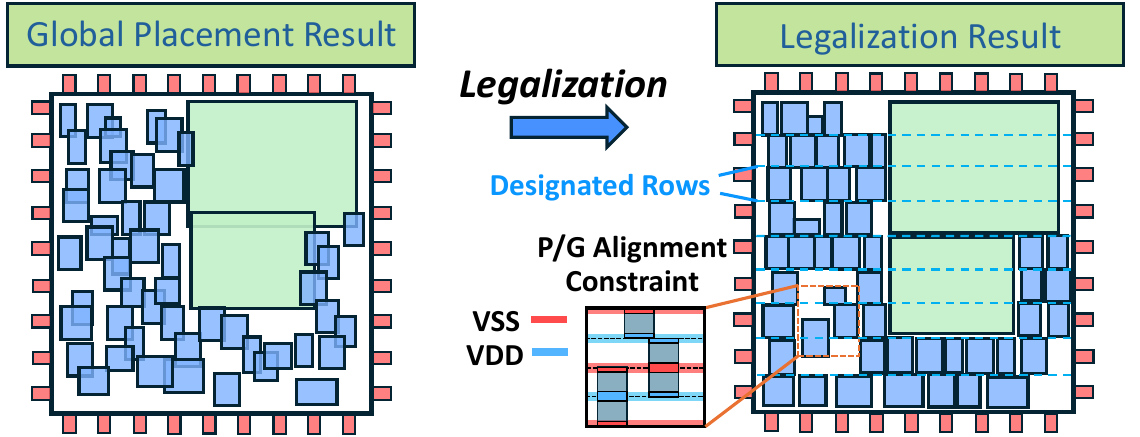}
\vspace{-10pt}
\caption{Example of layouts before and after legalization.}
\label{fig_legalization_introduction}
\vspace{-12pt}
\end{figure}

As electronic design automation (EDA) algorithms confront escalating demands from sub-10nm technology nodes~\cite{liRoutabilityDrivenFenceAwareLegalization2018} and billion-component designs, legalization~\cite{legalizing2004,Jason2008,NGLIC2024} emerges as a critical yet computationally intensive step in VLSI physical design. This stage, as shown in Fig.~\ref{fig_legalization_introduction}, transforms globally placed circuits~\cite{GlobalPlacementTCAD} into legally aligned layouts by resolving overlaps while maintaining placement quality—a prerequisite for successful routing. The proliferation of mixed-cell-height designs~\cite{wangEffectiveLegalizationAlgorithm2017,zhuMixedCellHeightLegalizationConsidering2020,daravICCAD2017CADContest2017}, where single-row and multi-row cells coexist to optimize performance and design flexibility, further complicates the problem: cell interactions propagate displacement across rows~\cite{chenNBLGRobustLegalizer2022}, while stringent design rules severely constrain solution spaces~\cite{liPinAccessibleLegalizationMixedCellHeight2022,liRoutabilityDrivenFenceAwareLegalization2018,chenMixedCellHeightPlacementDraintoDrain2022,liAnalyticalMixedCellHeightLegalization2019}. To address these challenges, existing legalization strategies~\cite{chenNBLGRobustLegalizer2022,chenOptimalLegalizationMixedcellheight2017,chenRobustModulusBasedMatrix2020,chowLegalizationAlgorithmMultiplerow2016,doFenceRegionAwareMixedHeightStandard2019,hungMixedCellHeightStandardCell2017,liAnalyticalMixedCellHeightLegalization2019,liPinAccessibleLegalizationMixedCellHeight2022,liRoutabilityDrivenFenceAwareLegalization2018,zhuMixedCellHeightLegalizationConsidering2020,zhouRobustNewtontypeIteration2023,zhouAcceleratedModulusbasedMatrix2023,wangEffectiveLegalizationAlgorithm2017} primarily follow two paradigms: purely analytical and heuristic-analytical-mixed approaches.

%%%Figure legalization introduction

%%%Figure introduction overview
\begin{figure*} [t]
\centering
\includegraphics[width=\linewidth]{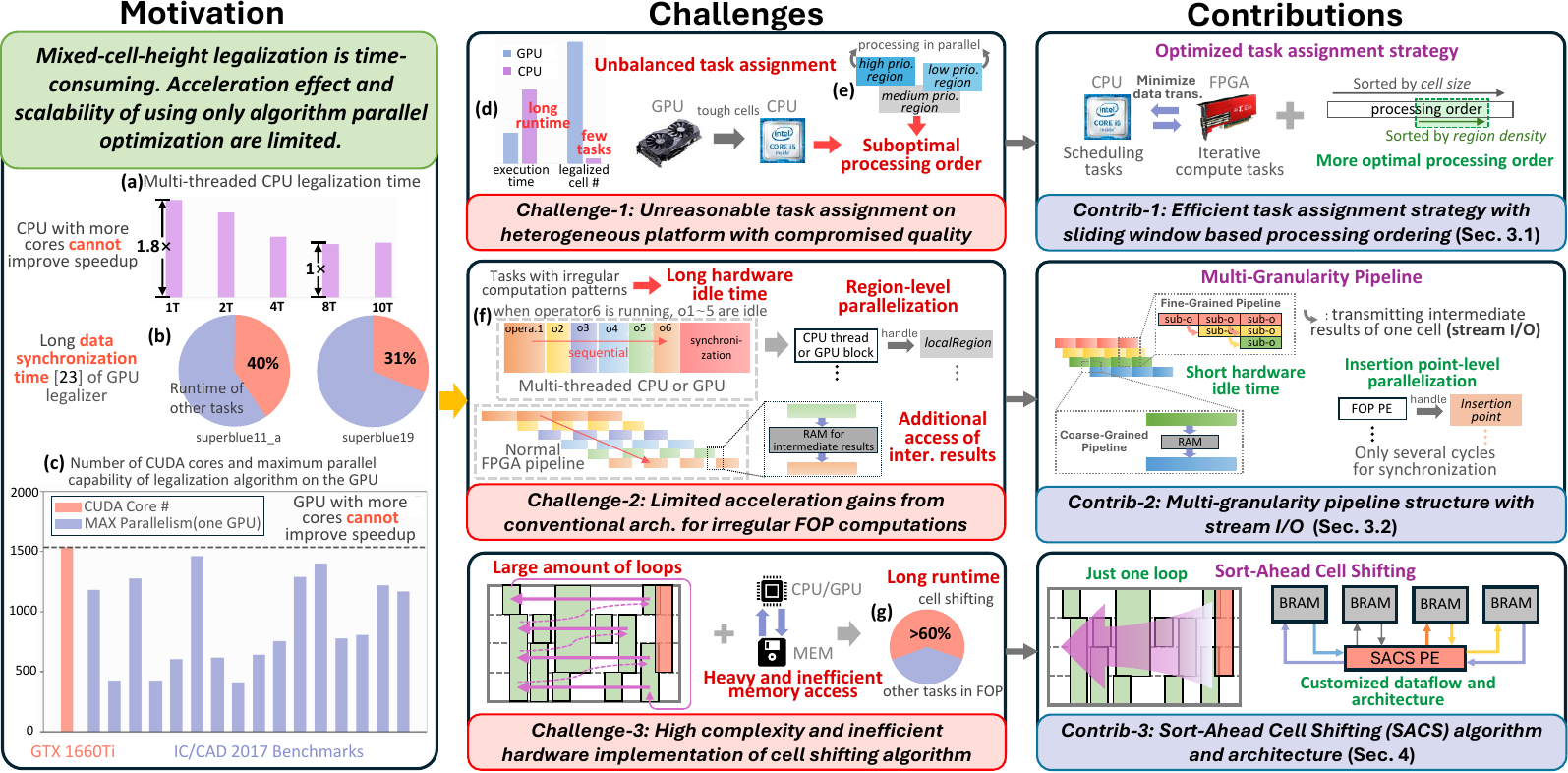}
\vspace{-21pt}
\caption{Overview of the challenges and contributions.}
\label{fig_intro_overview}
\vspace{-15pt}
\end{figure*}

%%% 简要介绍related work 引出challenge
Purely analytical methods~\cite{chenOptimalLegalizationMixedcellheight2017,liAnalyticalMixedCellHeightLegalization2019,chenRobustModulusBasedMatrix2020,chenNBLGRobustLegalizer2022} formulate legalization as common optimization problems such as quadratic programming, which can be solved by existing solvers. While theoretically capable of achieving optimal solutions, these methods suffer from high runtime and poor scalability, limiting their practicality. \iffalse For instance, processing a design with 900,000 cells takes over 27 times longer than processing one with 110,000 cells of similar density, even without accounting for additional constraints~\cite{chenNBLGRobustLegalizer2022}. This renders analytical methods impractical for large-scale designs.\fi To improve efficiency, these problems are simplified into smaller, faster-to-solve tasks, while compromising solution quality. 

In contrast, heuristic-analytical-mixed methods achieve a balance between runtime and solution quality by employing heuristic algorithms for the broader problem and applying analytical methods to smaller sub-problems for refinement. Heuristic components, typically based on greedy strategies augmented by manually designed constraints, enable faster convergence to locally optimal solutions. Among these, the Multi-row Global Legalization (MGL) algorithm~\cite{liPinAccessibleLegalizationMixedCellHeight2022} resolves overlaps and meets technology constraints efficiently, leveraging analytical methods for fine optimization. While outperforming many purely analytical methods~\cite{liPinAccessibleLegalizationMixedCellHeight2022}, its runtime remains a bottleneck for large-scale designs~\cite{yangMixedCellHeightLegalizationCPUGPU2022}, as the computational intensity of finding optimal placement position for cells. Moreover, iterative nature of VLSI physical design may require thousands of legalization steps, significantly impacting overall efficiency.

Despite efforts to enhance legalization quality, accelerating this time-consuming step presents underexplored challenges. Existing parallelization methods exhibit poor scalability and diminishing returns: CPU-based multi-threading (processing several unlegalized cells \iffalse non-overlapping regions \fi concurrently~\cite{liPinAccessibleLegalizationMixedCellHeight2022}) saturates at 8 threads (Fig.~\ref{fig_intro_overview}(a)), while GPU implementations~\cite{yangMixedCellHeightLegalizationCPUGPU2022} suffer from two inherent problems. First, coarse-grained parallelism leads to high synchronization overhead (Fig.~\ref{fig_intro_overview}(b)), causing GPU methods to underperform even CPU baselines. Second, the number of parallelizable regions falls short of available CUDA cores (Fig.~\ref{fig_intro_overview}(c)), which means using a GPU with more cores is inefficient. \iffalse making additional cores inefficient\fi These observations reveal a critical gap in current acceleration methods: neither conventional CPU multi-threading nor brute-force GPU parallelization effectively harness modern hardware for scalable legalization.

To further enhance acceleration and scalability, several persistent challenges must be addressed:

\textbf{Challenge-1}: Unbalanced task assignment on heterogeneous platforms limits speedup and degrades quality. For example, in~\cite{yangMixedCellHeightLegalizationCPUGPU2022}, tough cells are assigned to CPU, leading to CPU's long runtime with minimal task completion on it (Fig.~\ref{fig_intro_overview}(d)). This also disrupts the processing order of regions, significantly affecting legalization quality. Additionally, assigning many non-overlapping regions to GPU for parallel processing results in a suboptimal processing order (Fig.~\ref{fig_intro_overview}(e)), further compromising the overall legalization quality. 

\textbf{Challenge-2}: Finding optimal placement position (FOP), the bottleneck of MGL algorithm, exhibits irregular computational patterns~\cite{Phloem}, which conventional architectures struggle to handle efficiently. While multi-threaded CPU or GPU acceleration improves throughput, it leads to resource underutilization due to idle hardware during sequential operation execution (Fig.~\ref{fig_intro_overview}(f)). Similarly, normal FPGA pipelines incur inefficiencies, as operators must wait for predecessors to finish, causing idle time and delays from memory access when intermediate results are stored and retrieved. 

\textbf{Challenge-3}: High-complexity cell shifting dominates over 60\% of FOP runtime (Fig.~\ref{fig_intro_overview}(g)) due to two factors: 1) unpredictable number of full-region cell data traversals, and 2) hardware-agnostic algorithm design with unoptimized memory access patterns. Existing implementations~\cite{yangMixedCellHeightLegalizationCPUGPU2022,liPinAccessibleLegalizationMixedCellHeight2022} lack architecture-aware optimizations, leaving parallelism potential unexploited.

To overcome these challenges, we propose FLEX, a heterogeneous accelerator for mixed-cell-height legalization based on a CPU-FPGA platform. Our contributions are summarized as follows:

\vspace{-2pt}
\begin{itemize}
\item To the best of our knowledge, FLEX is the first CPU-FPGA-based accelerator designed for mixed-cell-height legalization, featuring co-optimization of algorithm and architecture.

\begin{figure*} [t]
\centering
\includegraphics[width=\linewidth]{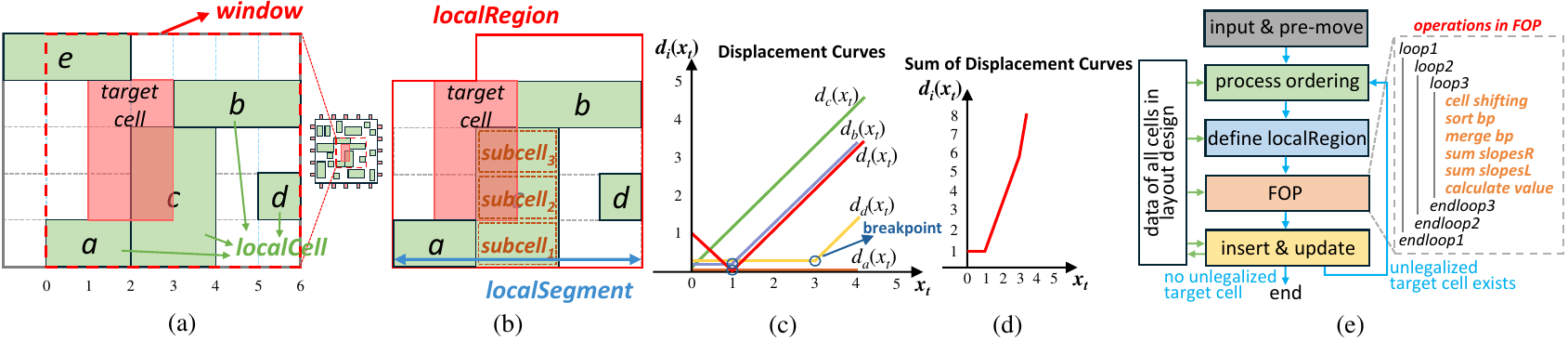}
\vspace{-22pt}
\caption{(a) Example of window and localCell. (b) Example of localRegion, localSegment, and subcell. (c) Example of displacement curves. (d) Example of the sum of displacement curves. (e) The legalization flow of using MGL algorithm~\cite{liPinAccessibleLegalizationMixedCellHeight2022}.}
\label{fig_background}
\vspace{-13pt}
\end{figure*}

\item FLEX applies an efficient task assignment strategy to minimize bandwidth usage between CPU and FPGA, significantly reducing runtime. Moreover, a cell processing ordering method is developed to enhance legalization quality. 

\item The most time-intensive FOP stage is accelerated through a multi-granularity pipeline design, achieving higher hardware efficiency and faster performance. 

\item For the loop-intensive cell shifting inside FOP, we propose a Sort-Ahead Cell Shifting algorithm to lower computational complexity and implement a high-performance architecture.

\end{itemize}

FLEX achieves up to 18.3$\times$ and 5.4$\times$ speedups over SOTA CPU-GPU~\cite{yangMixedCellHeightLegalizationCPUGPU2022} and multi-threaded CPU~\cite{liPinAccessibleLegalizationMixedCellHeight2022} legalizers with better scalability, while improving solution quality by 4\% and 1\%.

\vspace{-3pt}
\section{BACKGROUND} \label{section2}

In this section, we provide an overview of mixed-cell-height legalization problem, introduce key terms of the MGL algorithm~\cite{liRoutabilityDrivenFenceAwareLegalization2018}, and summarize the legalization flow. 

\vspace{-3pt}
\subsection{Mixed-Cell-Height Legalization Problem}

After global placement~\cite{GlobalPlacementTCAD}, the positions of $n$ standard cells are determined, each with an initial bottom-left coordinate ($x'_i$, $y'_i$), width $w_i$, and height $h_i$. The cell height is measured in units of standard row heights. Global placement ensures an optimal distribution of cells, so legalization must preserve this placement quality by minimizing the overall cell displacement~\cite{chowLegalizationAlgorithmMultiplerow2016}.

The problem is formally defined as follows: given a set of $m$ mixed-cell-height cells $C = \{c_1, c_2, ..., c_m\}$, each cell $c_i$ must be moved from its initial global placement position ($x'_i$, $y'_i$) to a legalized position ($x_i$, $y_i$) such that the displacement is minimized. The displacement is defined by Manhattan distance as:
\begin{equation}
\delta_i = \left| x_i - x'_i \right| + \left| y_i - y'_i \right| \label{delta},
\end{equation}
and the placement quality is typically measured by the average displacement, $S_{am}$~\cite{daravICCAD2017CADContest2017}:
\begin{equation}
S_{am} = \frac{1}{H}\sum_{h=1}^H \frac{1}{\left| C_h \right|} \sum_{c_i \in C_h} \delta_i \label{Sam},
\end{equation}
where $H$ is the largest cell height, and $C_h \subseteq C$ represents the set of all cells with height $h$.

Legalization also requires cells to align with placement sites and avoid overlaps. Each cell is further constrained by the chip’s power rail configuration, with boundaries defined as either Power or Ground (P/G). This imposes a P/G alignment constraint, \iffalse and conform to the chip's power rail configuration~\cite{chowLegalizationAlgorithmMultiplerow2016} (P/G alignment), \fi as illustrated in Fig.~\ref{fig_legalization_introduction}. Vertical movement is costly due to alignment constraints~\cite{chowLegalizationAlgorithmMultiplerow2016}, so the MGL algorithm restricts movement to the horizontal direction, simplifying the problem while maintaining placement quality.

\vspace{-5pt}
\subsection{Terminology of the MGL Algorithm} \label{section2_2}

To facilitate analysis and improvement of MGL algorithm in subsequent sections, we briefly introduce its key terms and concepts.

\vspace{-2pt}
\subsubsection{LocalSegment, LocalCell, and LocalRegion}\ 

The legalization problem is localized within a rectangular window \textit{W}, as shown in Fig.~\ref{fig_background}(a). For each row within \textit{W}, the longest continuous sequence of unblocked placement sites is defined as a \textit{localSegment}~\cite{chowLegalizationAlgorithmMultiplerow2016}. A cell is classified as a \textit{localCell} if it is entirely contained within the localSegments of \textit{W}. Together, all localSegments and localCells in \textit{W} form the \textit{localRegion}. The legalization of an unplaced target cell is simplified as a task to insert it into its localRegion and legalize all localCells. Additionally, a \textit{subcell} represents a one-row-height unit of a multi-row-height localCell within a localSegment. The number of subcells in a localCell equals to its height in row units. For example, in Fig.~\ref{fig_background}(b), cell \textit{c} spans three rows and thus contains three subcells.

\vspace{-2pt}
\subsubsection{Insertion Interval and Insertion Point}\ 

An insertion interval is the gap between two adjacent cells in a localSegment and an insertion point is a combination of insertion intervals across multiple rows. For example, the three intervals marked in Fig.~\ref{fig_SACS_compare3}(a) can be combined to get an insertion point for a three-row-height target cell.

\vspace{-2pt}
\subsubsection{Displacement Curve and Breakpoint}\ 

Even within a valid insertion point, the exact placement position of target cell can vary, resulting in different total displacements. \textit{Displacement curves} and \textit{breakpoints} are used to evaluate all possible placement positions within an insertion point. As shown in Fig.~\ref{fig_background}(c), a localCell's displacement curve is a piecewise linear function of target cell's placement position~\cite{chowLegalizationAlgorithmMultiplerow2016}. The turning points of these curves are referred to as breakpoints~\cite{liRoutabilityDrivenFenceAwareLegalization2018}. To identify the optimal placement position, all displacement curves are summed, as illustrated in Fig.~\ref{fig_background}(d). The x-coordinate with the minimum total displacement value is selected as the best.

\vspace{-3pt}
\subsection{Legalization Algorithm Flow}

Like existing legalization accelerators~\cite{yangMixedCellHeightLegalizationCPUGPU2022,liPinAccessibleLegalizationMixedCellHeight2022,liRoutabilityDrivenFenceAwareLegalization2018}, FLEX is also based on the MGL algorithm. The legalization flow is outlined in Fig.~\ref{fig_background}(e) and consists of five steps. 
\begin{enumerate}
\item[a)] \textbf{input \& pre-move}: Load the global placement results and temporarily position cells in the nearest designated rows while tolerating initial overlaps.
\item[b)] \textbf{process ordering}: Identify unlegalized target cells and determine the processing order.
\item[c)] \textbf{define localRegion}: Identify localSegments and localCells in currently processing target cell's window, and compute the localRegion density.
\item[d)] \textbf{FOP}: Employ triple-loop traversal to evaluate all candidate positions for minimal displacement.
\item[e)] \textbf{insert \& update}: Insert the target cell into the localRegion and update affected cells.
\end{enumerate}
Steps b) to e) are repeated until achieving complete legalization with zero overlaps and design rule violations.

FOP process, the most computationally intensive step of MGL, involves three nested loops traversing all insertion points in a localRegion. Each localRegion may contain hundreds of insertion points, making operations inside the inner loop of FOP the primary bottleneck of MGL. As illustrated in Fig.~\ref{fig_background}(e), operations inside the loop of FOP aim to identify the position with the minimum total displacement by evaluating all candidate positions. Key components of FOP include:

\begin{itemize}
\item \textbf{cell shift}: Resolve overlaps caused by target cell insertion through left-move (for cells on the left of target cell) and right-move (for cells on the right of target cell) phases.
\item \textbf{sort bp}: Gather and sort all breakpoints~(bp) by their x-coordinates.
\item \textbf{merge bp}: Traverse the sorted breakpoints, merging those with identical x-coordinates by accumulating their displacement curves' left-slopes and right-slopes.
\item \textbf{sum slopesR}: Forward traverse the merged breakpoints to compute cumulative right-slopes for all breakpoints to the left of \textit{bp}$_i$, storing the results in \textit{slopesR}$[i]$.
\item \textbf{sum slopesL}: Backward traverse the merged breakpoints to calculate cumulative left-slopes for breakpoints to the right of \textit{bp}$_j$, storing values in \textit{slopesL}$[j]$.
\item \textbf{calculate value}: Traverse the merged breakpoints to compute \textit{slopes} between each adjacent merged breakpoints by accumulating all displacement curves obtained by summing corresponding items in \textit{slopesR} and \textit{slopesL}. Then compute displacement value of each merged breakpoint based on \textit{slopes}, ultimately determining the position with the minimum displacement. This step and previous four steps are actually used to compute the sum of all displacement curves, as shown in Fig.~\ref{fig_background}(c) and (d).
\end{itemize}

\begin{figure}[t]
\centering
\includegraphics[width=85mm]{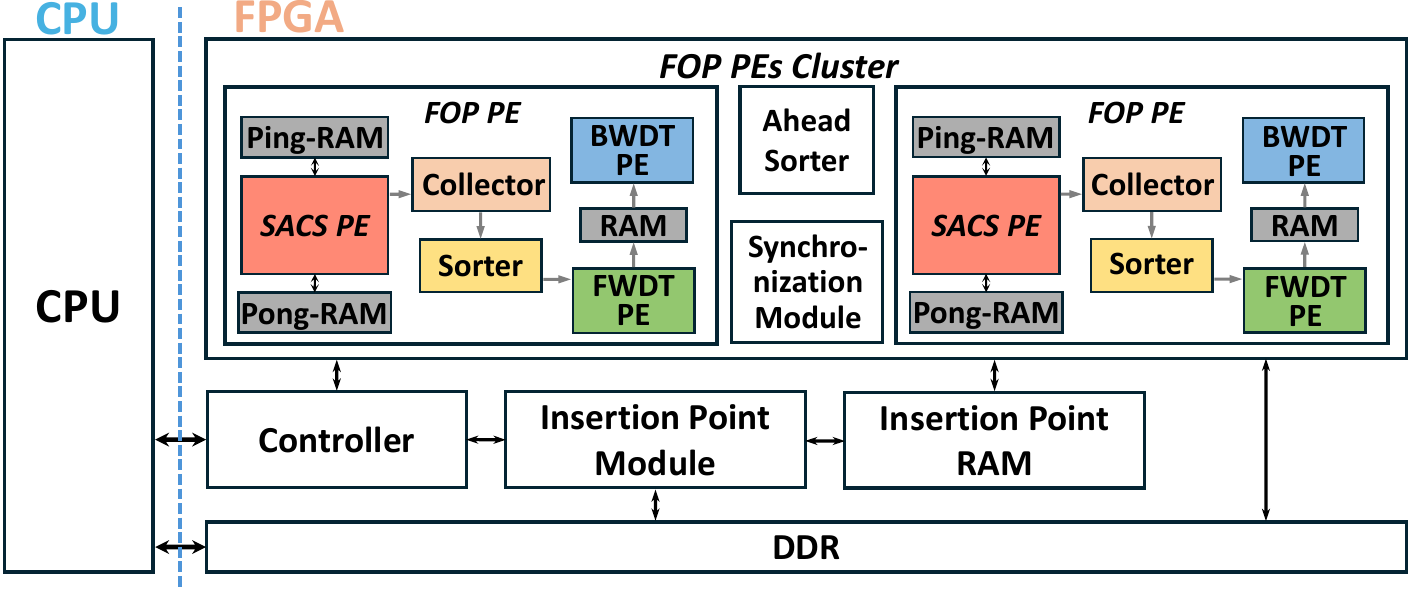}
\vspace{-22pt}
\caption{Architecture overview of FLEX.}
\label{fig_acceleration_framework}
\vspace{-15pt}
\end{figure}

\vspace{-3pt}
\section{HARDWARE-DRIVEN LEGALIZATION FLOW} \label{section3}

The architecture overview of FLEX is presented in Fig.~\ref{fig_acceleration_framework}. FLEX is specifically designed to accelerate the legalization flow depicted in Fig.~\ref{fig_background}(e), leveraging an FPGA-CPU platform. An controller manages the execution and data transfer across FPGA modules. The Insertion Point Module and RAM are utilized for computing and storing information pertaining to each insertion point. The FOP processing elements (PEs) cluster includes two FOP PEs, an Ahead Sorter, and a Synchronization Module. This section focuses on task assignment strategy and pipeline implementation within the FOP PE. 

\vspace{-3pt}

\subsection{Task Assignment Strategy and Processing Ordering}
\subsubsection{Task Assignment Between CPU and FPGA}\label{3_1} \ 

To address \textbf{Challenge-1}, we propose an efficient task assignment strategy that leverages the strengths of CPUs and FPGAs while minimizing bandwidth usage.

The first three steps in Fig.~\ref{fig_background}(e) are handled by the CPU with its flexibility and scheduling capabilities. Specifically: step a) is inherently serial~\cite{Ripple} and benefits from CPU's high clock speed; step b) involves dynamically scheduling, making it well-suited for CPU's flexibility. Although step c) could theoretically be processed in parallel~\cite{yangMixedCellHeightLegalizationCPUGPU2022}, we keep it on CPU for two reasons: (1) it accounts for only 3\% of total runtime on CPU, making FPGA acceleration unnecessary and power-inefficient; (2) the localRegion density computed in step c) is a critical metric for processing ordering in step b). Keeping both on CPU minimizes data transmission overhead.

Step d), the FOP process, including iterative and irregular computations, is divided into sub-operations (for one cell, not all cells in a region) and offloaded to FPGA with customized design. On the one hand, a fine-grained pipeline is designed to ensure high hardware utilization, leveraging FPGA's pipelining and parallelism nature. On the other hand, FPGA's on chip RAM is flexibly configured to meet the need for fast and large-scale memory access. 

Finally, step e), which involves a cell-shifting operation similar to step d), is kept on CPU despite its potential for FPGA acceleration. This is because step e) involves updating the positions of all moved cells, and assigning it to the FPGA would require transmitting all updated positions back to the CPU, introducing additional latency and interfering with steps b) and c). By keeping step e) on CPU, we eliminate unnecessary data transfer overhead, allowing the CPU to run at full capacity alongside the FPGA, reducing idle time for both. Effect of the assignment strategy is evaluated in Sec.~\ref{section5_3}.

\begin{figure*}[ht]
\centering
\includegraphics[width=\linewidth]{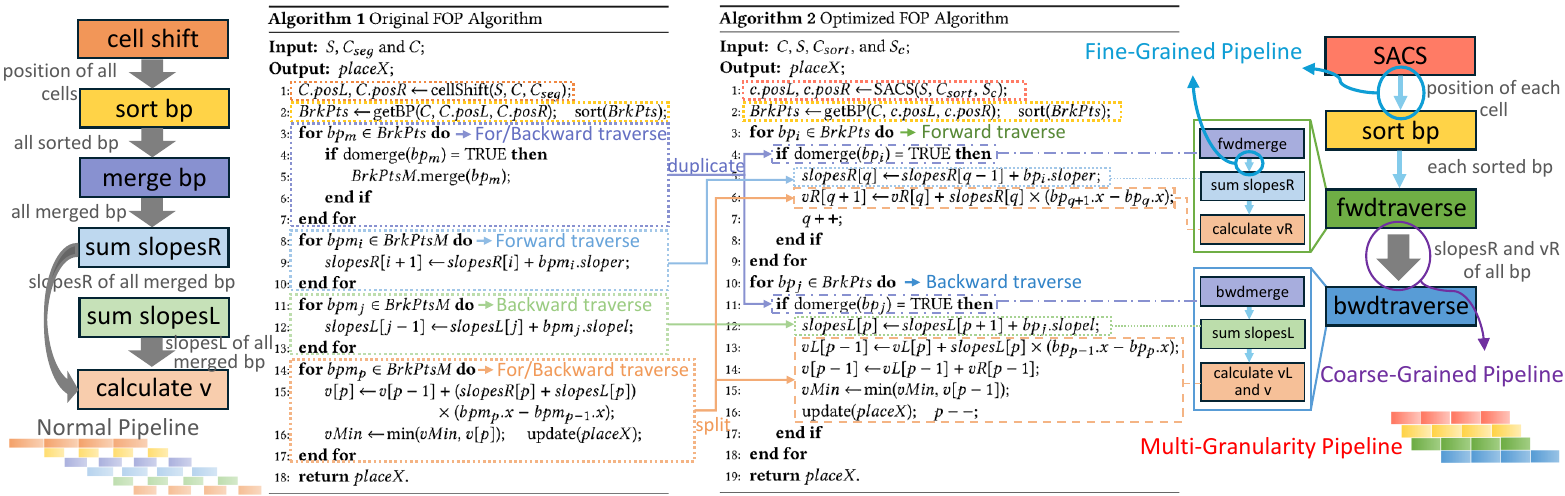}
\vspace{-20pt}
\caption{Comparison of original and optimized FOP algorithms.}
\label{fig_multi_granularity_pipeline}
\vspace{-15pt}
\end{figure*}

\subsubsection{Sliding Window Based Target Cell Processing Ordering} \

The processing order of target cells significantly influences the quality of heuristic legalization methods~\cite{chowLegalizationAlgorithmMultiplerow2016,liRoutabilityDrivenFenceAwareLegalization2018}. A widely adopted strategy prioritizes larger cells~\cite{yangMixedCellHeightLegalizationCPUGPU2022}, as smaller cells are easier to place and less likely to disturb neighboring cells. However, existing methods overlook the impact of localRegion density: in high-density regions, placing a cell can cause greater displacement to nearby cells. Additionally, parallel processing of multiple non-overlapping regions~\cite{yangMixedCellHeightLegalizationCPUGPU2022,liPinAccessibleLegalizationMixedCellHeight2022} may prematurely handle low-priority regions, deviating from the optimal processing order (Fig.~\ref{fig_intro_overview}(e)).

To address these limitations, we propose a method that considers both cell size and region density. Target cells are first sorted by size in descending order to generate an initial processing order sequence $S$. A sliding window $W_s$ is then applied, moving sequentially through $S$. The first cell in $W_s$, denoted as $C_{cur}$, is always processed first. While $C_{cur}$ is being processed, the next cell, $C_{next}$, is fixed, and other cells in $W_s$ are dynamically reordered based on their localRegions' density in descending order, ensuring a density-first priority for the remaining cells in $W_s$. Once $C_{cur}$ is processed, $W_s$ slides forward, making $C_{next}$ the new $C_{cur}$. To reduce runtime, if the localRegion of $C_{next}$ does not overlap with that of $C_{cur}$, its data will be preloaded into the free Ping-Pong RAM.

\subsection{Multi-granularity Pipeline for FOP}\label{section3_3}

\subsubsection{Motivation and Insights.} \

Traditional FPGA pipelines used to accelerate FOP, shown as the Normal Pipeline in Fig.~\ref{fig_intro_overview} (\textbf{Challenge-2}), suffer from inefficiencies. Each operation must wait for its predecessor to finish before starting, causing substantial idle time due to FOP's irregular computational patterns. Furthermore, intermediate results are stored in memory after each operation, incurring extra memory access delays when subsequent operators retrieve these results.

An ideal fine-grained pipeline would allow operators to output intermediate results as soon as sub-operations are done, enabling subsequent ones to begin execution without waiting for the completion of the entire predecessor operation. This requires a streaming input/output (stream I/O) model, where operators incrementally process input data and output results. However, the original FOP algorithms lack this capability. To address \textbf{Challenge-2}, we optimize FOP process for multi-granularity pipelining with two key improvements. First, we propose the Sort-Ahead Cell Shifting algorithm and architecture to handle the most complex cell-shifting operation, detailed in Sec.~\ref{section4}. Second, we restructure the last four operations in FOP to enhance pipelining efficiency.

\vspace{-4pt}
\subsubsection{Operation Reorganization for Bidirectional Traversals}\ 

Fig.~\ref{fig_multi_granularity_pipeline} compares the original and optimized FOP workflows. In the original approach~\cite{Ripple}, each operation must wait for its predecessor to finish, requiring additional RAM for intermediate data storage. In contrast, the optimized workflow (Fig.~\ref{fig_multi_granularity_pipeline} Algorithm 2) introduces stream I/O across most operations, except for the bidirectional traversals. Thin blue arrows indicate the immediate transfer of intermediate results for individual cells or breakpoints between operations. 

Enabling stream I/O requires not only optimized cell-shifting but also restructuring of the last four operations in the original FOP, which involve traversing breakpoints. Since \textit{sum slopesR} and \textit{sum slopesL} require bidirectional traversal of breakpoints, fine-grained pipelining between them is impractical because the output of one cannot be immediately utilized by the other. 

To achieve speedup, we implement coarse-grained pipelining between the two bidirectional traversals and fine-grained pipelining within operations traversing in the same direction by splitting and reorganizing them. Specifically, the merging step is duplicated into \textit{forward-merge} (\textit{fwdmerge}) and \textit{backward-merge} (\textit{bwdmerge}), while \textit{calculate v} is split into \textit{calculate vR}, \textit{vL}, and \textit{v}. As shown in Fig.~\ref{fig_multi_granularity_pipeline}, the original four operations are reorganized into two streamlined operations: \textit{fwdtraverse} and \textit{bwdtraverse}, based on the traversal order. Within \textit{fwdtraverse}, operations \textit{fwdmerge}, \textit{sum} \textit{slopesR}, and \textit{calculate vR} are executed sequentially in forward traverse order. Similarly, within \textit{bwdtraverse}, operations \textit{bwdmerge}, \textit{sum slopesL}, \textit{calculate vL}, and \textit{v} are executed in backward traverse order. Stream I/O is enabled between internal operations within \textit{fwdtraverse} and \textit{bwdtraverse}, allowing for streaming input to \textit{fwdtraverse} and streaming output from \textit{bwdtraverse} without introducing additional loops. This combination of fine-grained and coarse-grained pipelining forms the multi-granularity pipeline structure.

\begin{figure*} [t]
\centering
\includegraphics[width=162mm]{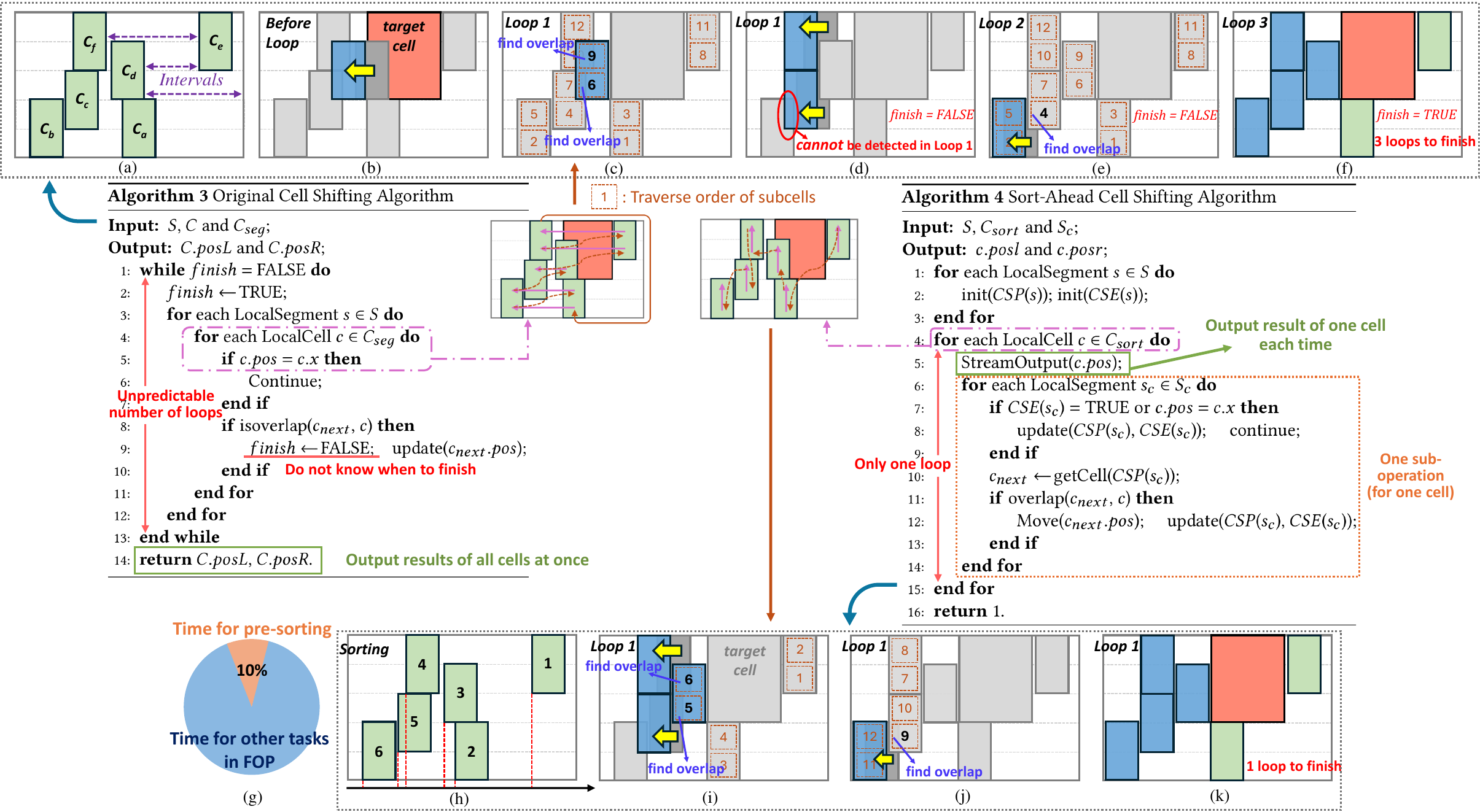}
\vspace{-12pt}
\caption{Comparison of the original cell shifting algorithm and the sort-ahead cell shifting algorithm.}
\label{fig_SACS_compare3}
\vspace{-14pt}
\end{figure*}

\vspace{-6pt}
\section{SORT-AHEAD CELL SHIFTING} \label{section4}

\subsection{Motivation}

As illustrated in Fig.~\ref{fig_intro_overview} \textbf{Challenge-3}, cell shifting is the most complex and time-consuming task within FOP, presenting significant problems for acceleration. Existing methods~\cite{Ripple,yangMixedCellHeightLegalizationCPUGPU2022} suffer from unpredictable multi-pass loops, making stream I/O and fine-grained pipelining infeasible. To address these issues, we propose the Sort-Ahead Cell Shifting (SACS) algorithm and its customized hardware architecture. SACS is integrated seamlessly into the multi-granularity pipelining framework for FOP, eliminating overlap resolution inefficiencies and enabling streaming outputs. 

\vspace{-6pt}
\subsection{Sort-Ahead Cell Shifting Algorithm}

Original cell shifting algorithm~\cite{liPinAccessibleLegalizationMixedCellHeight2022} (Fig.~\ref{fig_SACS_compare3} Algorithm 3) employs a \textit{finish} flag to indicate whether all overlaps are resolved in the current iteration. If overlaps persist, the loop continues iteratively until no cells are moved in the final pass. For example, in the left-move phase shown in Fig.~\ref{fig_SACS_compare3}(a)-(f), overlaps are resolved iteratively by traversing subcells. The traversal order—bottom to top inter-row and right to left intra-row—can result in new overlaps remaining undetected in the same iteration (Fig.~\ref{fig_SACS_compare3}(d)). Consequently, multiple iterations (three in this example) are required to resolve all overlaps, with each loop traversing all subcells sequentially (1 to 12).

In contrast, our proposed SACS algorithm (Fig.~\ref{fig_SACS_compare3} Algorithm 4) resolves overlaps in a single loop. Before shifting, all localCells are sorted by their x-coordinates (Fig.~\ref{fig_SACS_compare3}(h)). Cells are then processed sequentially: right to left for the left-move phase, and left to right for the right-move phase. Within each cell, subcells are processed from bottom to top. For instance, overlaps caused by subcell\textsubscript{5} and subcell\textsubscript{6} are resolved immediately by moving adjacent cells (Fig.~\ref{fig_SACS_compare3}(i)). Unlike the original algorithm, which leaves subsequent overlaps unresolved in the same pass, SACS detects and resolves all remaining overlaps during the current iteration (Fig.~\ref{fig_SACS_compare3}(i)-(j)). This allows the left-move process to complete in a single loop, compared to three loops required by the original algorithm.

In Fig.~\ref{fig_SACS_compare3} Algorithm 4, $S$ denotes the set of localSegments in a localRegion, and $S_c$ represents the subset of localSegments containing a specific localCell $c$, used for identifying adjacent cells for overlap detection. The SACS algorithm pre-sorts localCells into $C_{sort}$ by x-coordinates, ensuring predictable and sequential processing. This eliminates the need for multi-pass loops and ensures consistent overlap detection. Adjacent cells are identified using $CurSegPtr$ (CSP) and $CurSegEnd$ (CSE) structures: CSP tracks index of the next cell to process in each localSegment (row); CSE determines whether all cells in a localSegment have been processed.

Depending on left-move or right-move phase, CSP is initialized to the rightmost or leftmost cell in each localSegment. From line 4 to 15 in Algorithm 4, the outer loop iterates through all sorted cells in $C_{sort}$, while the inner loop processes their corresponding segments. Taking left-move as an example, since all cells to the right of the currently traversed cell $c$ have already been processed, no other cells will affect its position. Thus, its $pos$ can be directly output, as shown in line 5. Then it checks if all cells in each segment of $S_c$ have been processed or if $c$ has not yet moved. If no left-adjacent cell exists or no overlap occurs, CSP and CSE for segments of $S_c$ are updated. If a left-adjacent cell exists in segment $s$ and $c$ has moved, resulting in an overlap in $s$, the next cell in $s$ is processed to resolve it, followed by updating CSP and CSE. Unlike the original algorithm, SACS outputs each cell’s final position during the loop, enabling fine-grained pipelining between cell shifting and $sort$ $bp$.

\begin{figure*}[t]
\centering
\includegraphics[width=160mm]{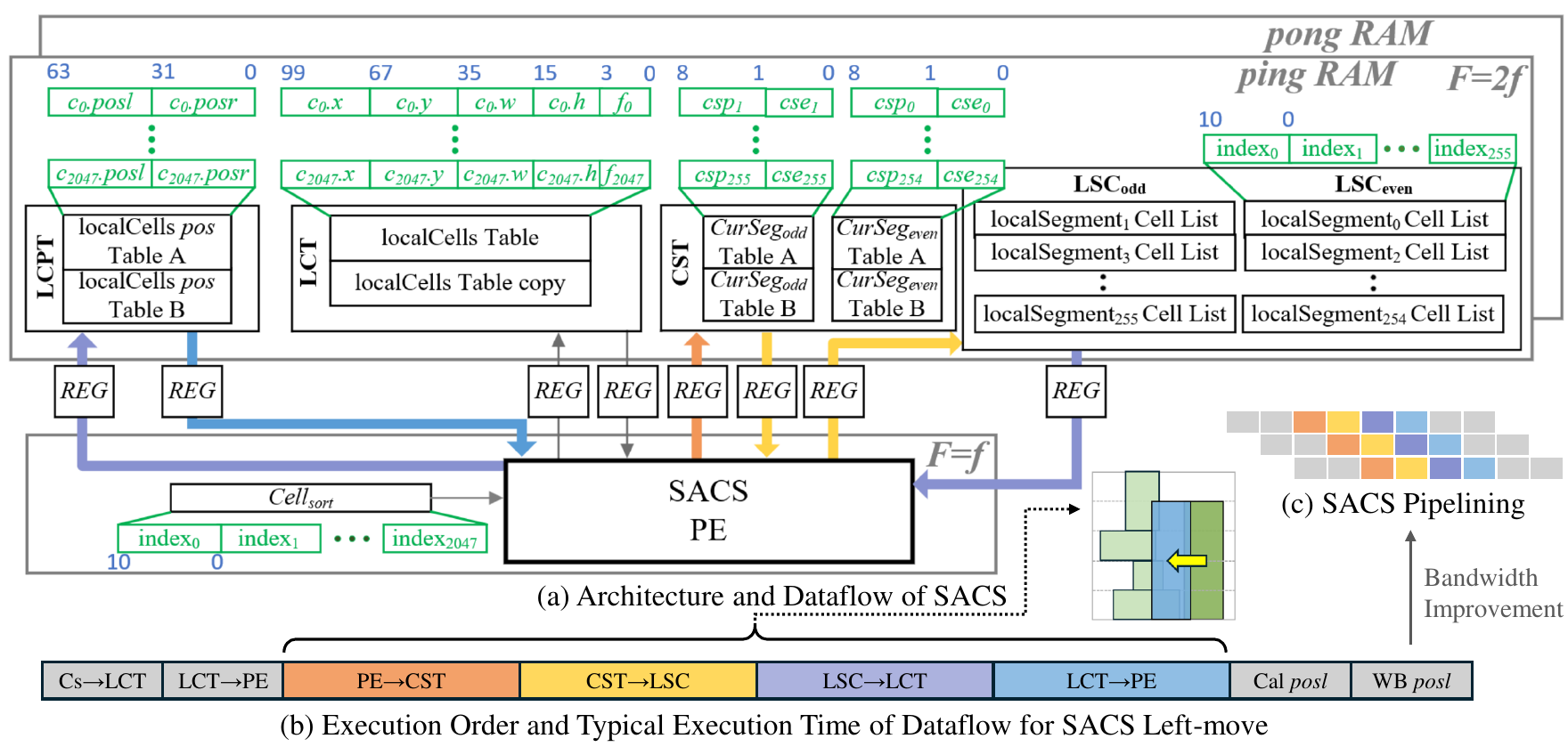}
\vspace{-13pt}
\caption{SACS architecture and pipeline for left-move phase.}
\label{fig_SACS_architecture_pipeline}
\vspace{-12pt}
\end{figure*}

\vspace{-8pt}
\subsection{Architecture for SACS Algorithm}
\subsubsection{Dataflow and Architecture}\ 

The SACS algorithm requires extensive data to represent the spatial relationships, such as cell positions, heights, widths, and associations with localSegments. While software implementations use complex data structures, FPGA-based accelerators require optimized memory designs. To this end, we introduce two tables: \textit{localCells Table} (LCT) stores fixed features and is updated only when processing a new localRegion; \textit{localCells pos Table} (LCPT) stores the recently updated positions, modified during runtime. Additionally, \textit{$Cell_{sort}$} (Cs) and \textit{localSegment Cell Lists} (LSC) store only cell indices to reduce memory usage, while the \textit{CurSeg Table} (CST) manages CSP and CSE. All memory structures shown in Fig.~\ref{fig_SACS_architecture_pipeline}(a) are implemented using FPGA BRAM for efficient execution.

Figure~\ref{fig_SACS_architecture_pipeline}(b) illustrates the dataflow and execution order for the SACS left-move phase. The dataflow of each stage is as follows:

\vspace{-2pt}
\begin{enumerate}
\item[a.] Cs$\xrightarrow{}$LCT: The next cell to be processed (\textit{Rcell}), is fetched from Cs, and its index is used to read features from LCT.
\item[b.] LCT$\xrightarrow{}$PE: Features of Rcell are loaded into the PE, which identifies the corresponding localSegments.
\item[c.] PE$\xrightarrow{}$CST: PE queries CST to retrieve adjacent cells and check segment completion. Since Rcell may belong to multiple localSegments, due to multi-row-heights, multiple addresses are queried in CST. Thicker arrows in Fig.~\ref{fig_SACS_architecture_pipeline}(a) represent data transfers for multiple cells. BRAM bandwidth limitations may increase latency during these queries ( Fig.~\ref{fig_SACS_architecture_pipeline}(b)).
\item[d.] CST$\xrightarrow{}$LSC: CSP retrieves the left-adjacent cell (\textit{Lcell}) from LSC for each localSegment.
\item[e.] LSC$\xrightarrow{}$LCT: Features of Lcells are fetched from LCT.
\item[f.] LCT$\xrightarrow{}$PE: Features of all Lcells are loaded into PE.
\item[g.] Cal \textit{posl}: PE calculates new \textit{posl} for Lcells to resolve overlaps.
\item[h.] WB \textit{posl}: Updated positions are written back to LCPT.
\end{enumerate}
\vspace{-3pt}

To reduce latency and hardware consumption, pre-sorting of localCells combines insertion~\cite{AMDInsertSort} and merge sorters~\cite{AMDMergeSort}. As shown in Fig.~\ref{fig_SACS_compare3}(g), sorting accounts for only 10\% of the total FOP runtime, introducing acceptable overhead. Evaluation presented in Sec.~\ref{5_4} demonstrates that the sorter requires minimal hardware resources.

\vspace{-2pt}
\subsubsection{Bandwidth Improvement for Multi-row-height Cell Access}\label{ImpBW}\ 

BRAM bandwidth can become a bottleneck for designs with many multi-row-height cells. To mitigate this, we employ:
\vspace{-2pt}
\begin{itemize}
    \item \textbf{Odd-Even RAM}: Divides LSC and CST into odd and even rows to double bandwidth. For instance, assuming each BRAM has two read and two write ports, accessing four adjacent cells spanning odd and even rows now takes a single cycle instead of two.
    \item \textbf{Ping-Pong Buffers}: LCPT and CST are initialized in parallel using two alternating buffers (Tables A and B in Fig.~\ref{fig_SACS_architecture_pipeline}(a)). This hides initialization latency during region processing.
    \item \textbf{Multiple Clock Domains}: LCT, LCPT, CST, and LSC, which require simultaneous access to multiple cells, are placed in a clock domain running at twice the frequency of the SACS PE. Additional split registers synchronize read addresses from the low-frequency domain to high-frequency domain and merge registers store data read from the latter.
\end{itemize}
\vspace{-2pt}
Additionally, LCT bandwidth is doubled by duplicating its memory, as its data is not row-dependent. Combined, these optimizations significantly reduce the latency of accessing multi-row-height cells, with minimal BRAM usage overhead. This enables highly efficient pipelining of the SACS algorithm, as illustrated in Fig.~\ref{fig_SACS_architecture_pipeline}(c), fully leveraging the parallelism inherent inside cell shifting operations.

\begin{table*} [t]
	\caption{Result comparison with state-of-the-art legalizer on IC/CAD 2017 contest benchmarks}
    \vspace{-8pt}
    \centering
    \small 
	\label{table}
	\setlength{\tabcolsep}{2.5pt} %表格线加粗
	\renewcommand\arraystretch{1.06} %增加表格行距
	\begin{tabular}{|m{2.2cm}<{\centering}|m{0.8cm}<{\centering}|m{1.0cm}<{\centering}|m{0.8cm}<{\centering}|m{0.9cm}<{\centering}|m{0.8cm}<{\centering}|m{0.9cm}<{\centering}|m{0.8cm}<{\centering}|m{0.9cm}<{\centering}|m{0.8cm}<{\centering}|m{0.9cm}<{\centering}|m{0.8cm}<{\centering}|m{0.8cm}<{\centering}|m{0.8cm}<{\centering}|}
	%保持水平\垂直居中
		\hline 
		\multirow{2}*{Benchmarks} & \multirow{2}*{Cell \#} & \multirow{2}*{Den. (\%)} & \multicolumn{2}{c|}{TCAD'22-MGL} & \multicolumn{2}{c|}{DATE'22} & \multicolumn{2}{c|}{ISPD'25} & \multicolumn{5}{c|}{Ours} \\ 
        \cline{4-14}
        ~ & ~ & ~ & AveDis & Time(s) & AveDis & Time(s) & AveDis & Time(s) & AveDis & Time(s) & Acc(T) & Acc(D) & Acc(I)\\
        \hline
        \hline
        \renewcommand\arraystretch{1} %增加表格行距
		%Q & \multicolumn{8}{c|}{ middle} \\ \hline %合并水平单元格
		des\_perf\_1 & 112644 & 90.6 & 0.967 & 4.74 & 1.05 & 3.47 & 0.66 & 7.51 & 0.665 & 1.322 & 3.6$\times$ & 2.6$\times$ & 5.7$\times$ \\ 
		des\_perf\_a\_md1 & 108288 & 55.1 & 0.919 & 1.81 & 0.92 & 2.00 & 1.20 & 8.38 & 0.904 & 0.727 & 2.5$\times$ & 2.8$\times$ & 11.5$\times$ \\ 
		des\_perf\_a\_md2 & 108288 & 55.9 & 1.148 & 1.67 & 1.32 & 2.00 & 1.12 & 16.64 & 1.144 & 0.663 & 2.5$\times$ & 3.0$\times$ & 25.1$\times$ \\ 
		des\_perf\_b\_md1 & 112644 & 55.0 & 0.675 & 1.28 & 0.70 & 6.85 & 0.65 & 20.34 & 0.635 & 0.375 & 3.4$\times$ & \textbf{18.3$\times$} & 54.2$\times$ \\ 
        des\_perf\_b\_md2 & 112644 & 64.7 & 0.618 & 1.31 & 0.72 & 1.75 & 0.70 & 1.11 & 0.653 & 0.501 & 2.6$\times$ & 3.5$\times$ & 2.2$\times$ \\ 
        edit\_dist\_1\_md1 & 130661 & 67.4 & 0.664 & 0.98 & 0.67 & 1.67 & 0.63 & 2.68 & 0.646 & 0.347 & 2.8$\times$ & 4.8$\times$ & 7.7$\times$ \\ 
        edit\_dist\_a\_md2 & 127413 & 59.4 & 0.614 & 1.30 & 0.73 & 1.80 & 0.67 & 2.22 & 0.650 & 0.547 & 2.4$\times$ & 3.3$\times$ & 4.1$\times$ \\ 
        edit\_dist\_a\_md3 & 127413 & 57.2 & 0.783 & 1.78 & 0.91 & 3.92 & 0.79 & 19.21 & 0.771 & 0.897 & 2.0$\times$ & 4.4$\times$ & 21.4$\times$ \\ 
        fft\_2\_md2 & 32281 & 82.7 & 0.721 & 0.29 & 0.68 & 0.45 & 0.68 & 1.74 & 0.694 & 0.112 & 2.6$\times$ & 4.0$\times$ & 15.5$\times$ \\ 
        fft\_a\_md2 & 30625 & 32.3 & 0.563 & 0.22 & 0.65 & 0.32 & 0.75 & 0.51 & 0.604 & 0.041 & \textbf{5.4$\times$} & 7.8$\times$ & 12.4$\times$ \\ 
        fft\_a\_md3 & 30625 & 31.2 & 0.531 & 0.15 & 0.56 & 0.34 & 0.59 & 0.39 & 0.567 & 0.036 & 4.2$\times$ & 9.4$\times$ & 10.8$\times$ \\ 
        pci\_b\_a\_md1 & 29517 & 49.5 & 0.652 & 0.33 & 0.63 & 0.58 & 0.92 & 0.70 & 0.699 & 0.106 & 3.1$\times$ & 5.5$\times$ & 6.6$\times$ \\ 
        pci\_b\_a\_md2 & 29517 & 57.7 & 0.839 & 0.47 & 0.91 & 0.62 & 0.85 & 2.12 & 0.838 & 0.130 & 3.6$\times$ & 4.8$\times$ & 16.3$\times$ \\ 
        pci\_b\_b\_md1 & 28914 & 26.6 & 0.781 & 0.31 & 0.48 & 0.62 & 1.14 & 0.88 & 0.821 & 0.085 & 3.6$\times$ & 7.3$\times$ & 10.4$\times$ \\ 
        pci\_b\_b\_md2 & 28914 & 18.3 & 0.704 & 0.32 & 0.63 & 0.45 & 1.01 & 1.69 & 0.746 & 0.072 & 4.4$\times$ & 6.3$\times$ & 23.5$\times$ \\ 
        pci\_b\_b\_md3 & 28914 & 22.2 & 0.925 & 0.34 & 0.87 & 0.45 & 1.09 & 1.92 & 0.945 & 0.082 & 4.1$\times$ & 5.5$\times$ & 23.4$\times$ \\ 
        \hline 
        \hline
        \multicolumn{3}{|c|}{Average} & 0.757 & 1.08 & 0.78 & 1.71 & 0.84 & 5.50 & 0.749& 0.378 & 2.9$\times$ & 4.5$\times$ & 14.7$\times$  \\ \hline
        \multicolumn{3}{|c|}{Ratio} & 1.01 & 2.86 & 1.04 & 4.52 & 1.12 & 14.67 & 1.00 & 1.00 & - & - & - \\
        \hline
	\end{tabular}
	\label{tab1}
    \vspace{-6pt}
\end{table*}

Although the SACS algorithm can also be implemented on CPUs, the actual performance gains are limited. The memory access pattern remains irregular, and additional indexing structures like CSP and CSE incur software overhead. In contrast, our FPGA implementation customizes the on-chip RAM and memory dataflow to match the algorithm, enabling highly efficient pipelining and parallelism. Furthermore, FPGA-based sorting~\cite{AMDInsertSort,AMDMergeSort} is significantly faster and more scalable than CPU-based sorting, especially for sorting large amounts of data. Thus, SACS achieves far higher efficiency when deployed on FPGA.

\vspace{-5pt}
\section{EXPERIMENTAL RESULTS}\label{section5}
\vspace{-1pt}
\subsection{Configuration and Dataset}

FLEX is implemented on a system with an Intel Core i5 CPU, 32 GB RAM, and an AMD Alveo U50 FPGA running at 285 MHz. Comparisons are made against the CPU-GPU legalizer~\cite{yangMixedCellHeightLegalizationCPUGPU2022} (Intel Core i5 CPU, NVIDIA GeForce GTX 1660 Ti GPU), the multi-threaded CPU legalizer~\cite{liPinAccessibleLegalizationMixedCellHeight2022} (Intel Xeon CPU, 8 cores, 64 GB RAM), and the purely analytical GPU-accelerated legalizer~\cite{LEGALM} (NVIDIA A800 GPU). All experiments use the IC/CAD 2017 contest benchmarks~\cite{daravICCAD2017CADContest2017}, with baseline results taken directly from the respective papers.

\vspace{-8pt}
\subsection{Overall Performance}

The performance of FLEX is evaluated against two MGL-based heuristic legalizers and one pure analytical legalizer: the CPU-GPU-based legalizer~\cite{yangMixedCellHeightLegalizationCPUGPU2022}, the multi-threaded CPU legalizer~\cite{liPinAccessibleLegalizationMixedCellHeight2022} and GPU-accelerated LEGALM~\cite{LEGALM}. Table 1 provides a detailed comparison of results and the meaning of each column is: 

\begin{itemize}
    \item Benchmarks: IC/CAD 2017 contest test cases.
    \item Cell \#: Number of cells to be legalized.
    \item Den.(\%): Design density, calculated as total cell area divided by available free area.
    \item TCAD’22-MGL, DATE'22, ISPD'25: Results from~\cite{liPinAccessibleLegalizationMixedCellHeight2022} for MGL algorithm only, results from~\cite{yangMixedCellHeightLegalizationCPUGPU2022}, and results from~\cite{LEGALM}.
    \item AveDis: Average displacement of legalized cells, computed using Equation (2).
    \item Time(s): Runtime for the legalization task.
    \item Acc(T), Acc(D) and Acc(I): Speedup of this work compared to CPU legalizer~\cite{liPinAccessibleLegalizationMixedCellHeight2022}, CPU-GPU legalizer~\cite{yangMixedCellHeightLegalizationCPUGPU2022} and GPU-based analytical legalizer~\cite{LEGALM}.
\end{itemize}
From Table 1, FLEX achieves 2.6$\times$–18.3$\times$ speedup over the CPU-GPU legalizer and 2$\times$–5.4$\times$ over the multi-threaded CPU legalizer. \iffalse For a VLSI design that originally takes an average of 20 seconds to legalize and requires 1000 iterations of optimization, FLEX can reduce the runtime from 5.5 hours to less than one hour. \fi Additionally, FLEX enhances solution quality, particularly in high-density cases like “des\_perf\_1,” where FLEX achieves lower average displacement than the heuristic baselines. While the purely analytical legalizer from ISPD'25 is not our primary baseline, FLEX outperforms it in both runtime and legalization quality, demonstrating the superiority of the heuristic-analytical-mixed approach. 

\vspace{-3pt}
\subsection{Breakdown Analysis} \label{section5_3}

To understand the contributions of each optimization, we perform a breakdown analysis of FLEX's acceleration mechanisms.

\begin{figure} [t]
\centering
\includegraphics[width=76mm]{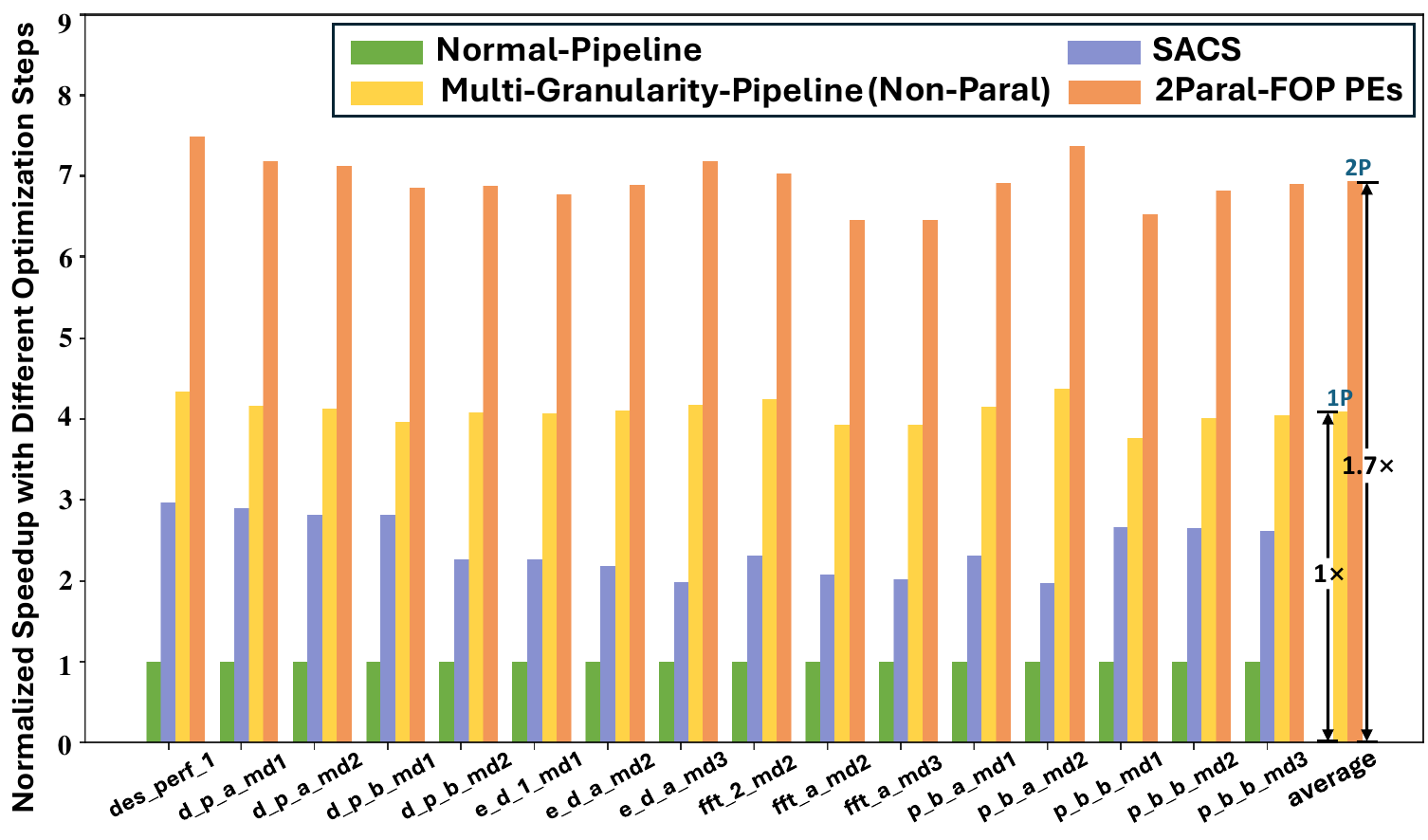}
\vspace{-13pt}
\caption{Speedup with different optimization steps on FPGA.}
\label{fig_normalized_speedup_4}
\vspace{-15pt}
\end{figure}

\textbf{FOP Optimization.} FLEX's FPGA-based FOP acceleration consists of three main components: cell shifting optimization (Sec.~\ref{section4}), multi-granularity pipelining (Sec.~\ref{section3_3}), and parallelization of FOP PE (shown in Fig.~\ref{fig_acceleration_framework}). Fig.~\ref{fig_normalized_speedup_4} shows the normalized speedup by these optimizations: cell shifting optimization (SACS) achieves 2$\times$–3$\times$, multi-granularity pipelining gains an additional speedup of 1$\times$–2$\times$, and 2-parallelism FOP PE results in a further speedup of 1.6$\times$–1.9$\times$.

\begin{figure}[t]
\centering
\includegraphics[width=77mm]{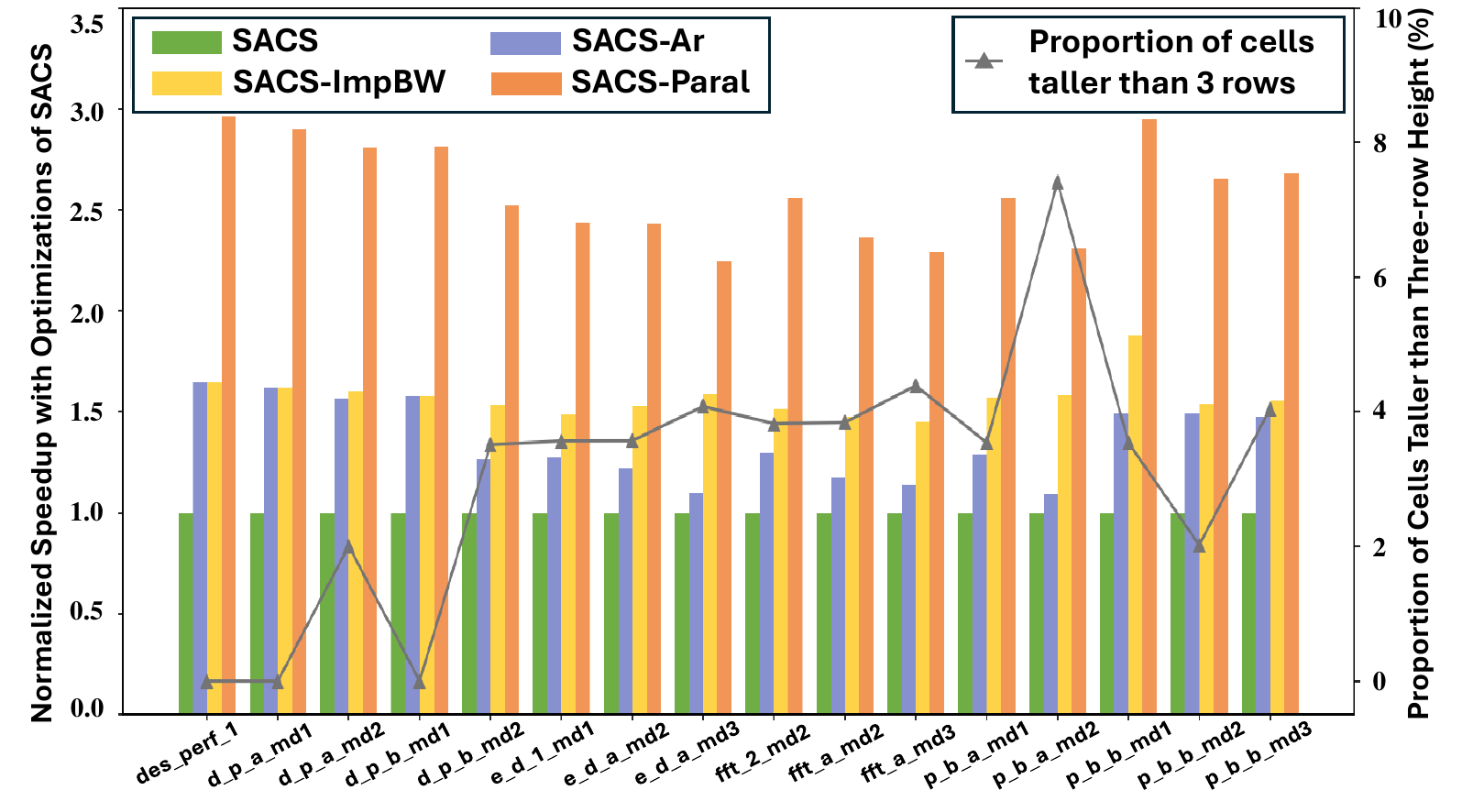}
\vspace{-15pt}
\caption{Speedup with different optimization steps of SACS and percentage of cells taller than three-row height.}
\label{fig_speedup_inside_SACS}
\vspace{-16pt}
\end{figure}

\textbf{SACS Optimization.} SACS optimizations include architecture improvements, bandwidth optimization, and parallelization of left and right moves. Fig.~\ref{fig_speedup_inside_SACS} illustrates the normalized speedup achieved through various optimizations. The gray line in Fig.~\ref{fig_speedup_inside_SACS} indicates the proportion of cells taller than three rows, which plays a critical role in determining the effectiveness of bandwidth optimization. Specifically, the speedup between “SACS-Ar” and “SACS-ImpBW” correlates with this ratio. A higher proportion of tall cells increases the likelihood of simultaneous access to multiple adjacent cells, thereby amplifying the benefits of bandwidth improvements. In benchmarks without cells taller than three rows (e.g., the 1st, 2nd, and 4th benchmarks), “SACS-ImpBW” demonstrates no additional speedup over “SACS-Ar.” However, in the “pci\_b\_a\_md2” benchmark, the high proportion of such tall cells results in a significant acceleration due to the bandwidth optimization.

%%%Figure Speedup of task assignment

%%%Figure speedup of optimization inside SACS 

\begin{figure} [t]
\centering
\includegraphics[width=77mm]{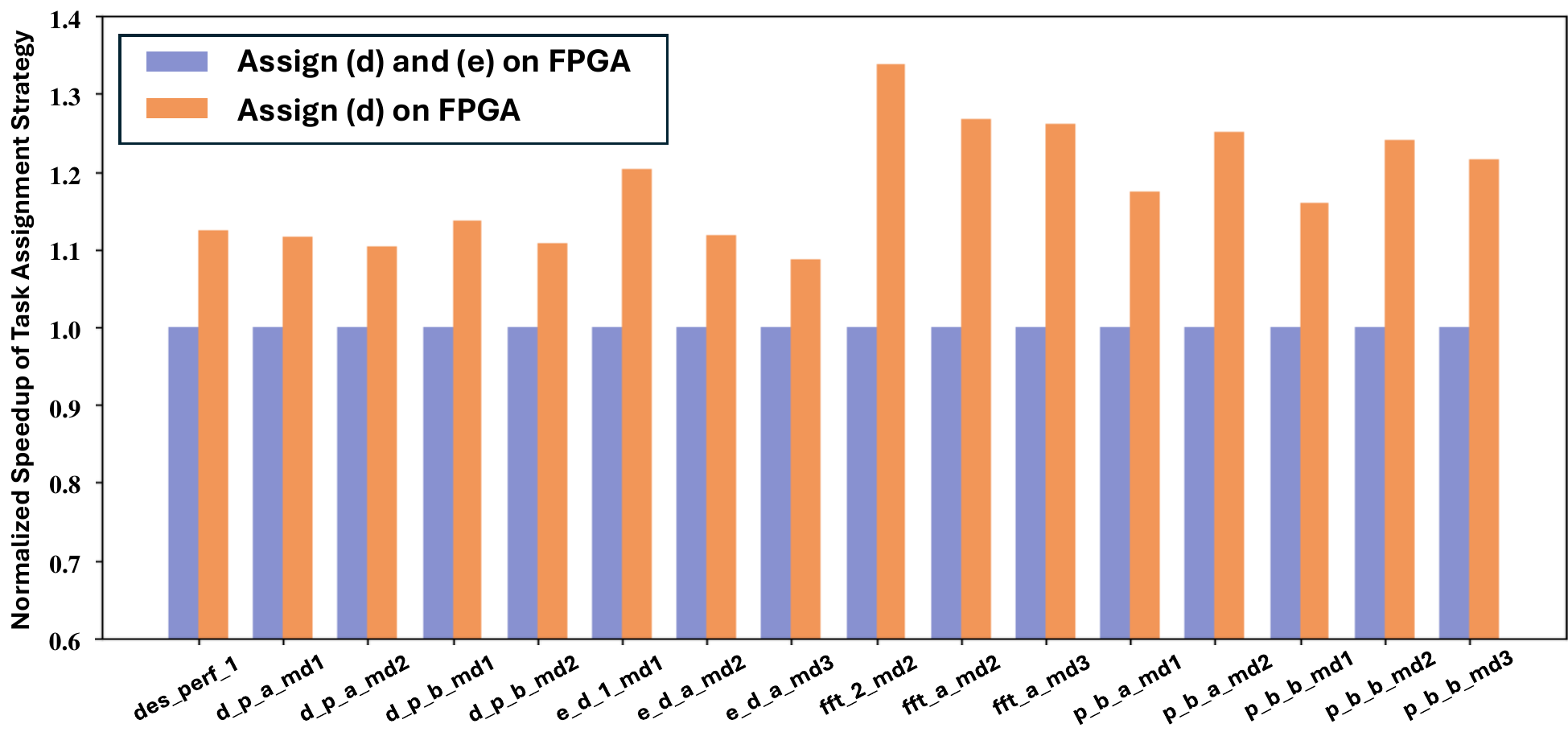}
\vspace{-12pt}
\caption{Speedup of proposed task assignment method.}
\label{fig_task_assign_speedup}
\vspace{-14pt}
\end{figure}

\textbf{Task Assignment Optimization.} To evaluate our task assignment strategy between CPU and FPGA, we compare different allocation schemes. As discussed in Sec.~\ref{3_1}, steps a), b) and c) are performed on the CPU due to algorithmic constraints. Thus, the comparison focuses on step d) being executed solely on FPGA versus both d) and e) being offloaded to FPGA. Fig.~\ref{fig_task_assign_speedup} shows that our method, which keeps step e) on the CPU, achieves an average speedup of 1.2$\times$ compared to offloading both steps to FPGA.

A large part of the speedup brought by task assignment optimization comes from the reduction of communication overhead between CPU and FPGA, as mentioned in Sec.~\ref{3_1}. In addition, our hardware design further reduces the impact of communication by enabling overlap between data transfer and computation. The employed Ping-Pong RAM structure (Fig.~\ref{fig_acceleration_framework}) and deeply pipelined architecture (Fig.~\ref{fig_multi_granularity_pipeline}) allow the FPGA to process one region while simultaneously loading data for the next. The processing ordering strategy also ensures that the required data for the next region ($C_{next}$) is preloaded in advance. In fact, the visible communication cost approximately equals to the initial data transfer time of the first processed region, which is negligible in the overall runtime.

\vspace{-2pt}
\subsection{Resource Utilization and Scalability} \label{5_4}

Table 2 details FPGA resource utilization. Two parallel FOP PEs can process two insertion points within the same localRegion simultaneously. Additionally, the sorter for each localRegion is not duplicated, resulting in a less than two times increase in \textit{LUT} and \textit{FF} usage when enabling two parallelism of PEs. Speedup can be further improved by increasing the number of FOP PEs while BRAM may become a resource bound, but this can be addressed by using URAM with a slight FPGA clock frequency penalty.

Regarding scalability, multi-threaded CPU legalizers exhibit performance saturation at 8 threads due to the limitations of thread-level parallelism (Fig.~\ref{fig_intro_overview}(a)). Similarly, while the scalability of the CPU-GPU legalizer is not explicitly analyzed in~\cite{yangMixedCellHeightLegalizationCPUGPU2022}, the experimental results in Fig.~\ref{fig_intro_overview}(b) and (c) indicate that its performance is constrained by significant data synchronization overhead. These methods rely on processing multiple localRegions in parallel, which necessitates extensive synchronization to update the positions of all moved cells, thereby limiting their scalability. 

By contrast, FLEX takes a different approach to scalability by avoiding region-level parallelism, which inherently reduces synchronization costs. For example, in the 2-parallelism FOP PE configuration, FLEX computes the displacements of two insertion points for the same target cell concurrently. Afterward, a simple synchronization operation is performed to compare the displacement results and select the smaller one, taking several clock cycles. This low-level parallelism minimizes synchronization overhead, allowing FLEX to scale more effectively. 

%%%Figure Speedup of task assignment

\begin{small}
\begin{table} [t]
	\caption{Hardware resource consumption on FPGA}
    \vspace{-8pt}
    \centering
	\label{table2}
	\setlength{\tabcolsep}{3pt} 
	\renewcommand\arraystretch{1.06} 
	\begin{tabular}{|m{3cm}<{\centering}|m{1cm}<{\centering}|m{1cm}<{\centering}|m{1cm}<{\centering}|m{1cm}<{\centering}|}
        \hline 
        ~ & LUTs & FFs & BRAMs & DSPs \\
        \hline
        No parallelism of FOP PE & 59837 & 67326 & 391 & 8 \\
        %\hline
        2 parallelism of FOP PE & 86632 & 91603 & 738 & 12 \\
        %\hline
        Available & 871680 & 1743360 & 1344 & 5952 \\
        \hline
	\end{tabular}
	\label{tab2}
    \vspace{-18pt}
\end{table}
\end{small}

Fig.~\ref{fig_normalized_speedup_4} demonstrates that 2-parallelism FOP PEs on FPGA achieves an average 1.7$\times$ speedup compared to the non-parallel setup, approaching near-linear scalability. As on-chip memory and FPGA resource capacities grow, FLEX can achieve even higher levels of parallelism, further enhancing its performance for large-scale tasks. In contrast, the multi-threaded CPU legalizer shows diminishing returns as the number of threads increases. Specifically, 2-threaded CPU runtime is reduced by only 20\% compared to a single-threaded implementation (Fig.~\ref{fig_intro_overview}(a)). With 8 threads, the speedup saturates, achieving only a 1.8$\times$ acceleration, falling short of ideal scalability.

\vspace{-3pt}
\section{RELATED WORKS}\label{section6}

\textbf{Legalization Algorithms:}  
Abacus~\cite{spindlerAbacusFastLegalization2008} is one of the most widely used algorithms for traditional single-cell-height standard cell legalization and has been integrated into several SOTA placers~\cite{liaoDREAMPlace40Timingdriven2022,guDREAMPlace30MultiElectrostatics2020,linDREAMPlaceDeepLearning2021}. It employs dynamic programming to achieve optimal solutions with minimal displacement for cells within a row. However, it is not well-suited for mixed-cell-height legalization tasks, as moving multi-row-height cells can introduce overlaps in adjacent rows. To address mixed-cell-height legalization, several works~\cite{chenOptimalLegalizationMixedcellheight2017,liAnalyticalMixedCellHeightLegalization2019,chenRobustModulusBasedMatrix2020} formulate the problem as a quadratic programming (QP) task, which is subsequently reformulated into a linear complementarity problem (LCP). However, to ensure convergence and equivalence between the QP and LCP formulations, additional operations are required~\cite{liAnalyticalMixedCellHeightLegalization2019}. Furthermore, these methods often leave some cells in illegal states for specific benchmarks, which can degrade the overall solution quality. Chen \textit{et al.}\cite{chenNBLGRobustLegalizer2022,open_NBLG} treats the legalization problem as a resource allocation task, employing a negotiation-based approach to achieve excellent displacement results. However, this method still suffers from pin shorts/access violations and edge spacing violations in several benchmarks\cite{liPinAccessibleLegalizationMixedCellHeight2022}. Chow \textit{et al.}\cite{chowLegalizationAlgorithmMultiplerow2016} proposes a multi-row local legalization (MLL) algorithm for mixed-cell-height legalization. After cells are pre-aligned to their nearest row and power rail positions, MLL evaluates all possible insertion points and selects the best position for each target cell. Building on this, Li \textit{et al.}\cite{liRoutabilityDrivenFenceAwareLegalization2018,liPinAccessibleLegalizationMixedCellHeight2022} introduces a multi-row global legalization (MGL) method to handle fence region and routability constraints. Unlike MLL, MGL accumulates the displacement between the global placement positions and final target positions, enabling identification of better insertion points for unlegalized cells.

\textbf{Accelerators for VLSI Legalization:}
Most EDA algorithm accelerators~\cite{TaroRTL,LUFRF,guoUltrafastCPUGPU2021,linRTLCUDAGPU2023} leverage GPUs for parallelism. For legalization, a CPU-GPU-based legalization accelerator~\cite{yangMixedCellHeightLegalizationCPUGPU2022} utilizes GPU parallelization to optimize tasks by exploiting different levels of concurrency. It processes non-overlapping regions and handling different rows to refine search windows in parallel. However, due to the lack of efficient queue data structure support on GPUs, a brute-force approach is used. This method employs multiple threads to evaluate single-row intervals and their validity in parallel, while simultaneously computing legalization costs and determining optimal insertion points. Although this approach achieves high parallelism, it incurs significant data synchronization overhead. Furthermore, the scheduler of it assigns tough cells to CPU, leaving GPU to handle only simpler tasks, which results in CPU spending significant time on those challenging cases. 

\vspace{-3pt}
\section{CONCLUSION}\label{section7}

We propose FLEX, an FPGA-CPU accelerator for mixed-cell-height legalization in VLSI physical design. Through efficient task assignment and architecture-algorithm co-optimization, FLEX achieves up to 18.3$\times$ and 5.4$\times$ speedups over SOTA GPU and CPU legalizers, while improving quality by 4\% and 1\% respectively. Our work demonstrates FPGAs' effectiveness for accelerating tasks with irregular computational patterns, suggesting potential for broader application across VLSI physical design as FPGA-based data centers become more prevalent.

\vspace{-3pt}
\begin{acks}
We would like to thank Dr. Haocheng Li and Prof. Evangeline F.Y. Young from the Chinese University of Hong Kong. This work was partially supported by the RGC GRF grant 16214123.
\end{acks}

\bibliographystyle{ACM-Reference-Format}
\bibliography{legalization_algorithm}
\end{document}